\documentclass[journal,a4paper,onecolumn]{IEEEtran}

\usepackage{tikz}
\usepackage{epsfig}
\usepackage{epstopdf}
\usepackage{siunitx}
\usepackage{pbox}
\usepackage{makecell}
\usepackage{multirow}
\usepackage{amssymb}
\usepackage{amsmath}
\usepackage[normalem]{ulem}

\usepackage{blindtext}
\usepackage{etoolbox}
 
\makeatletter

\docsvlist{\@oddhead,\@evenhead,\ps@headings,\ps@IEEEtitlepagestyle,\ps@IEEEpeerreviewcoverpagestyle}
\makeatother

\begin{document}


\title{A small-signal GFET equivalent circuit considering an explicit contribution of contact resistances}

\author{Anibal Pacheco-Sanchez, Javier N. Ramos-Silva, Eloy Ramírez-García, David Jiménez 

\thanks{This work has received funding from the European Union’s Horizon 2020 research and innovation programme under grant agreements No GrapheneCore2 785219 and No GrapheneCore3 881603, from Ministerio de Ciencia, Innovación y Universidades under grant agreement RTI 2018-097876-B-C21(MCIU/AEI/FEDER, UE). This  article  has been partially  funded  by  the European Regional Development Funds (ERDF)  allocated  to  the  Programa Operatiu FEDER de Catalunya 2014-2020, with the support of the Secretaria d’Universitats i Recerca of the Departament d’Empresa i Coneixement of the Generalitat de Catalunya for emerging technology clusters to  carry  out  valorization  and  transfer  of  research  results.  Reference  of  the  GraphCAT  project:  001-P-001702. This article has also received funding from Instituto Politécnico Nacional
under the contract no. SIP/20192062.  \newline \indent A. Pacheco-Sanchez and D. Jiménez are with the Departament d'Enginyeria Electr\`{o}nica, Escola d'Enginyeria, Universitat Aut\`{o}noma de Barcelona, Bellaterra 08193, Spain, e-mail: AnibalUriel.Pacheco@uab.cat \newline \indent J. N. Ramos-Silva and E. Ramírez-García are with Instituto Politécnico Nacional, UPALM, Edif. Z-4 3er Piso, Cd. de México, 07738, México.}
}
\maketitle
\makeatletter
\def\ps@IEEEtitlepagestyle{
  \def\@oddfoot{\mycopyrightnotice}
  \def\@evenfoot{}
}
\def\mycopyrightnotice{
  {\footnotesize
  \begin{minipage}{\textwidth}
  \centering
© 2020 IEEE.  Personal use of this material is permitted.  Permission from IEEE must be obtained for all other uses, in any current or future media, including reprinting/republishing this material for advertising or promotional purposes, creating new collective works, for resale or redistribution to servers or lists, or reuse of any copyrighted component of this work in other works.
  \end{minipage}
  }
}

\begin{abstract}
\boldmath
A small-signal equivalent circuit for graphene field-effect transistors is proposed considering the explicit contribution of effects at the metal-graphene interfaces by means of contact resistances. A methodology to separate the contact resistances from intrinsic parameters, obtained by a de-embedding process, and extrinsic parameters of the circuit is considered. The experimental high-frequency performance of three devices from two different GFET technologies is properly described by the proposed small-signal circuit. Some model parameters scale with the device footprint. The correct detachment of contact resistances from the internal transistor enables to assess their impact on the intrinsic cutoff frequency of the studied devices. 
\end{abstract}
%
%

\IEEEpeerreviewmaketitle
\section{Introduction}
\label{ch:intro}

Small-signal characterization of graphene field-effect transistors (GFETs) has enabled the demonstration and immediate parametric assessment at high-frequency (HF) of this emerging technology \cite{TanAnd14}-\cite{YuHe17_}. One of the key device parameters to be considered in GFETs is the contact resistance $R_{\rm{c}}$ related to bias-dependent potential barriers at the metal-graphene (MG) interfaces and additional interface layers \cite{GiuDiB17}-\cite{PacFei20}, i.e., $R_{\rm{c}}$ is the series combination of a bias-dependent resistance and a bias-independent resistance. The separation of these effects is not trivial \cite{ChaJim15}. In general, values of $R_{\rm{c}}$ can be obtained either by test structure characterization \cite{TanAnd14}, \cite{LyuLu16}, \cite{DenFad20}, by analytical models \cite{BonAsa19}, by \textit{S}-parameters measurements \cite{YuHe17}, \cite{PasWei17}, \cite{YuHe17_}, by \emph{I-V}-based extraction methods \cite{PacFei20} or by a fitting process \cite{SanXu18}. Regardless of the method to obtain it, in small-signal equivalent circuits (ECs) of GFETs, the extracted $R_{\rm{c}}$ has been usually either considered in the extrinsic part of the model only \cite{TanAnd14}, \cite{YuHe17}, \cite{SanXu18} or included in the intrinsic circuit \cite{LyuLu16}, \cite{PasWei17}, \cite{DenFad20}. It must be noticed that standard de-embedding procedures with test dummy patterns can not subtract the total effect of $R_{\rm{c}}$ from the intrinsic device. Hence, either intrinsic or extrinsic circuit elements compensate the impact of $R_{\rm{c}}$ by overestimating or underestimating other parameters depending on where $R_{\rm{c}}$ has been considered in the small-signal model.

The contribution from the source contact resistance $R_{\rm{sc}}$ and drain contact resistance $R_{\rm{dc}}$ has been generally embraced by $R_{\rm{c}}$ in a symmetrical manner, i.e., ${\textstyle R_{\rm{sc}} = R_{\rm{dc}} = R_{\rm{c}}/2}$ \cite{TanAnd14}, \cite{PasWei17}, \cite{BonAsa19}. This symmetrical distribution of effects is a good initial approximation but the real conditions can differ since the potential barrier at each MG interface, represented by $R_{\rm{sc/dc}}$, can vary depending on the operating bias point. In contrast to other approaches where $R_{\rm{C}}$ is lumped either in the extrinsic \cite{TanAnd14}, \cite{YuHe17}, \cite{SanXu18} or intrinsic \cite{LyuLu16}, \cite{PasWei17}, \cite{DenFad20} part of the EC, in this work a straightforward method considering asymmetric contact resistances in the EC separated from the intrinsic and extrinsic networks has been presented. The proposed approach has been validated with experimental data from two GFET technologies reported elsewhere \cite{YuHe17}, \cite{DenFad20}, \cite{YuHe17_}.

\section{Equivalent circuit and parameter extraction} \label{ch:DUT}

\begin{figure}[!htb]
\centering
\includegraphics[width=0.45\textwidth]{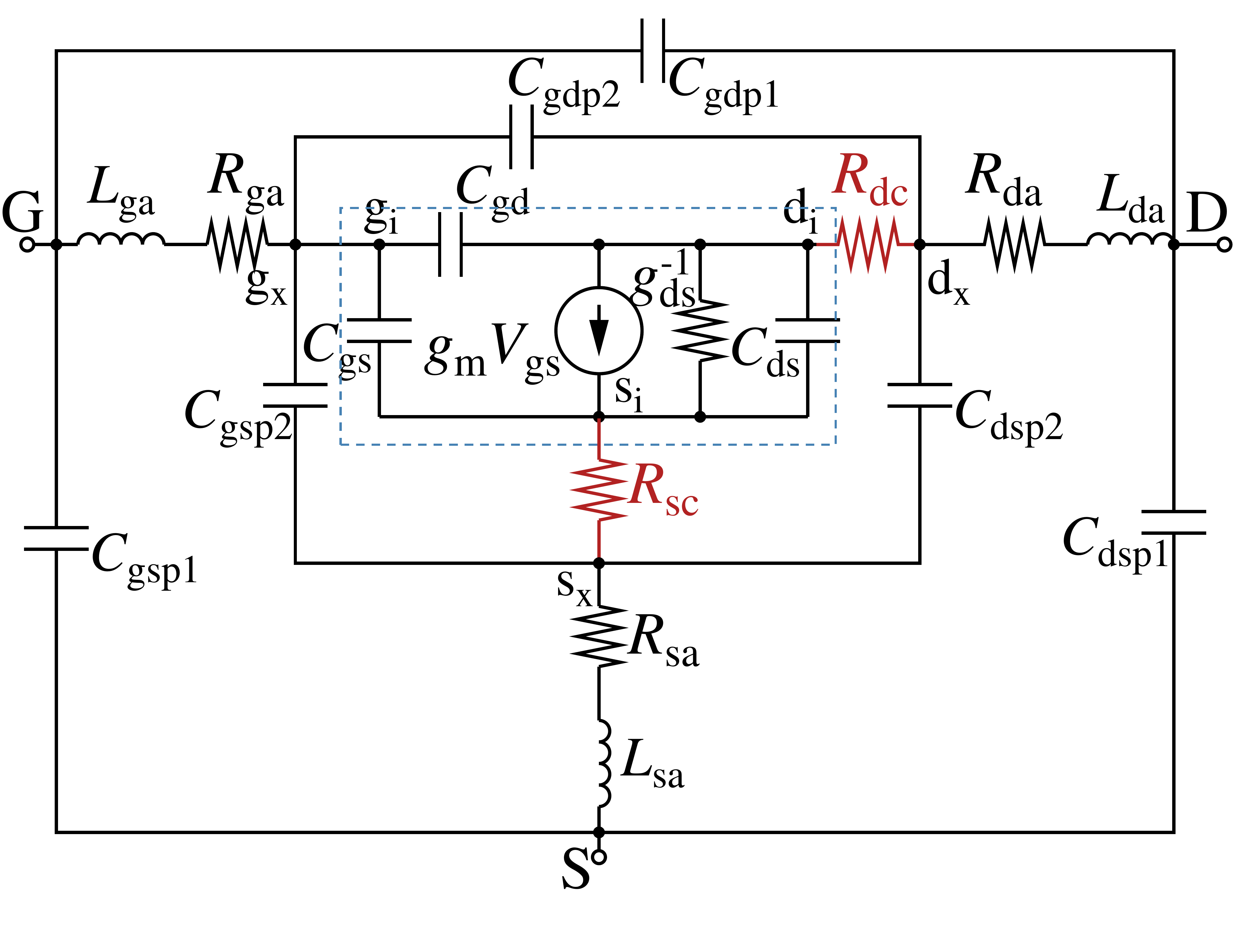}
\vspace{-0.4cm}\caption{Small-signal GFET equivalent circuit considering contact resistances effects related to MG interfaces. The intrinsic part is within the dashed box}
\label{fig:ckt}
\end{figure}

The small-signal EC considered in this work is shown in Fig. \ref{fig:ckt}. Non-negligible device contact resistances have been considered in this EC. This approach has been useful as well for describing the small-signal HF performance of other emerging transistors \cite{RamPac20} and it has been followed for the first time in GFETs here. The intrinsic elements are between the $\rm g_i$, $\rm s_i$ and $\rm d_i$ nodes. Source and drain contact resistances including the contribution of a potential barrier at the MG interfaces are between $\rm s_i$ and $\rm s_x$ nodes and $\rm d_i$ and $\rm d_x$ nodes, respectively. Extrinsic parasitic elements are between $\rm g_x$, $\rm s_x$ and $\rm d_x$ and the acces points of the device $\rm G$, $\rm S$ and $\rm D$, respectively. Conventional de-embedding techniques do not eliminate the impact of contact resistances on the intrinsic parameters. This effect should be removed in order to obtain an accurate description of the intrinsic device.

The parasitic effects of the metallic access lines and pads are represented by the gate/source/drain parasitic inductance $L_{\rm{ga/sa/da}}$, the gate/source/drain parasitic resistance $R_{\rm{ga/sa/da}}$ and the gate-to-source/gate-to-drain/drain-to-source parasitic capacitances $C_{\rm{gspx/gdpx/dspx}}$. The intrinsic transistor consists of the gate-to-source/gate-to-drain/drain-to-source capacitances $C_{\rm{gs/gd/ds}}$, the intrinsic output resistance $1/g_{\rm{ds}}$ and the voltage-controlled current defined by the intrinsic transconductance $g_{\rm{m}}$ and the intrinsic gate-to-source voltage $V_{\rm{gs}}$. In the context of graphene devices in general, the contact resistances $R_{\rm{sc/dc}}$ embrace both external and internal device phenomena \cite{GiuDiB17}-\cite{PacFei20} and their impact on the intrinsic device performance can not be separated by standard de-embedding procedures. The extraction of the small-signal parameters used here has been introduced elsewhere \cite{RamPac20} for a different low-dimensional transistor technology. This methodology is applied in this work to two GFET technologies and it is summarized as follows. 

$C_{\rm gsp1/gdp1/dsp1}$ can be obtained from the admittance matrix $[Y_{\rm{open}}]$ associated to the HF characterization of an open dummy structure using the same architecture and materials as the active device. Similarly, $L_{\rm ga/sa/da}$ and $R_{\rm ga/sa/da}$ are calculated from the HF impedance matrix $[Z_{\rm{short}}]$ of short dummy structures. The HF characterization of a pad dummy structure yields an admittance matrix $[Y_{\rm{pad}}]$ from which $C_{\rm gsp2/gdp2/dsp2}$ are obtained. The de-embedded device admittance matrix $[Y_{\rm{dem}}]$, including the impact of $R_{\rm{c}}$\footnote{$[Y_{\rm{dem}}]$ does not include the contribution of elements outside ${\textstyle\rm g_{\rm x}}$, ${\textstyle\rm d_{\rm x}}$ and ${\textstyle\rm s_{\rm x}}$ in the EC. $[Y_{\rm{dem}}]$ here is $[Y_{\rm INT}]$ in \cite{TieHav05} if ${\textstyle R_{\rm c}=0}$.}, is calculated with the raw extrinsic admittance matrix $[Y_{\rm{raw}}]$ and $[Y_{\rm{open}}]$, $[Z_{\rm{short}}]$ and $[Y_{\rm{pad}}]$ by following a three-step de-embedding method \cite{RamPac20}, \cite{TieHav05}. 

\begin{figure}[!htb]
\centering
\includegraphics[width=0.45\textwidth]{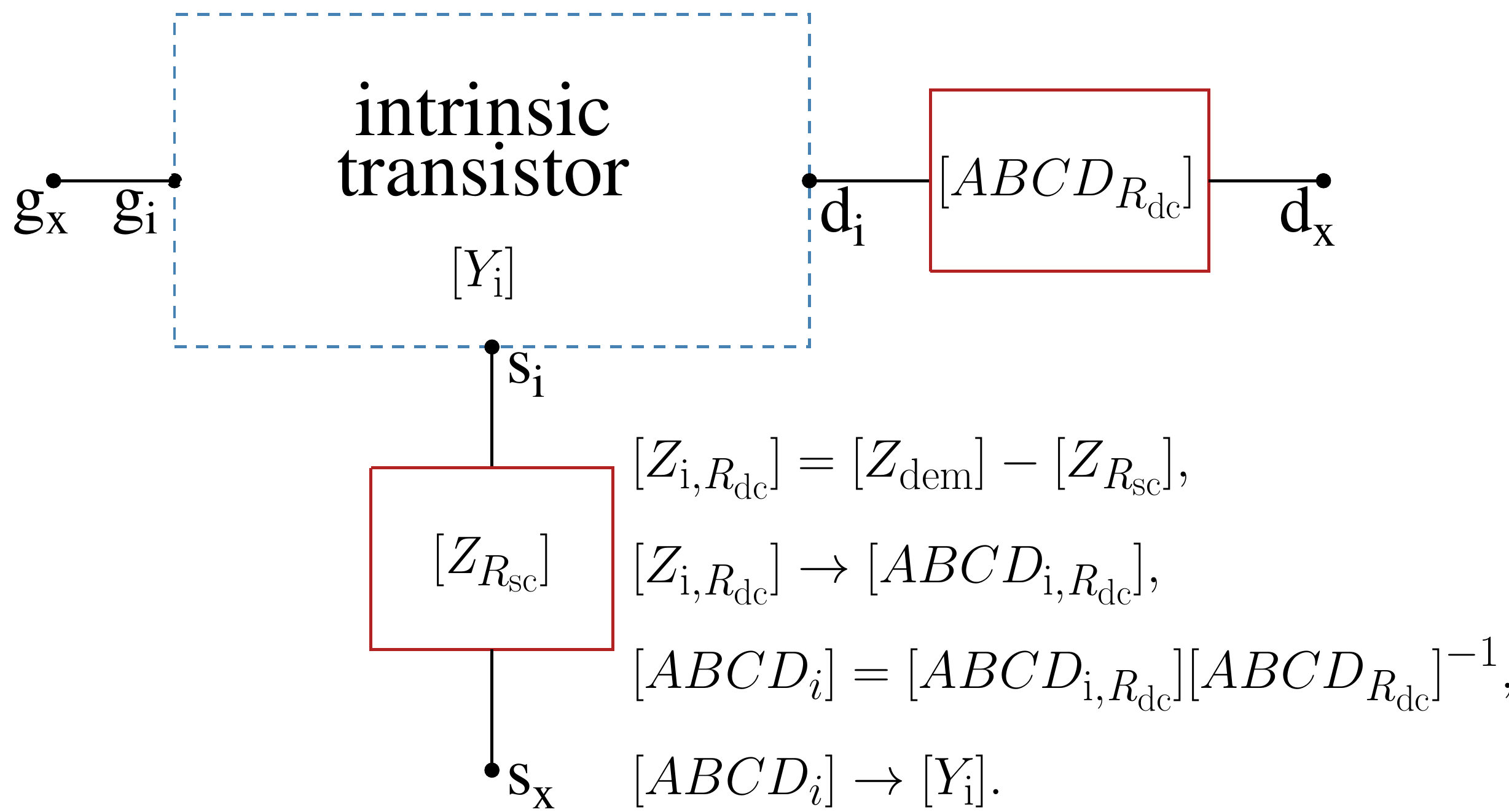}
\vspace{-0.3cm}\caption{Schematic representation and methodology to obtain intrinsic transistor characteristics without the impact of contact resistances as described in \cite{RamPac20}.}
\label{fig:ckt_int}
\end{figure}

In order to characterize correctly the intrinsic GFET, the contribution of the contact resistances should be removed from $[Y_{\rm{dem}}]$ since their impact has not been eliminated by the de-embedding methodology: the dummy structures do not include the potential barriers at the MG interfaces. This non-trivial approach is performed here, under the condition of linear matrices, by following the matrix algebra explained in \cite{RamPac20} after identifying the corresponding blocks and connection topologies of the intrinsic transistor and contact resistances as shown in Fig. \ref{fig:ckt_int}. This procedure is summarized as follows: the impedance matrix of the intrinsic device including only the contribution of the drain contact resistance $[Z_{\rm i,\mathit{R}_{\rm dc}}]$, obtained by removing the impedance matrix of $R_{\rm sc}$ from $[Z_{\rm dem}]$, has been transformed into its $[ABCD_{\rm i,\mathit{R}_{\rm dc}}]$ representation in order to obtain the intrinsic device matrix by multiplying it by the inverse $[ABCD]$-matrix of the remaining contact resistance, i.e., $[ABCD_{R_{\rm dc}}]^{-1}$. The intrinsic $[ABCD_{\rm i}]$ is then transformed into the intrinsic admittance matrix $[Y_{\rm{i}}]$.

A reliable calculation of each intrinsic element of the small-signal model is enabled by obtaining $[Y_{\rm{i}}]$. An initial value for $R_{\rm{sc}/dc}$, required for this procedure, can be set by considering a symmetrical disposal of $R_{\rm{c}}$, the value of which can obtained by one of the methods listed in Section \ref{ch:intro}. The asymmetric condition of $R_{\rm{c}}$ is achieved by an additional optimization step involving a correct description of experimental $S$-parameters. Notice that non-quasi static effects have not been considered in the model since their impact has been projected to be negligible at frequencies equal or lower than the cutoff frequency in GFETs, which is the frequency range analyzed here \cite{PasJim20}. 

The experimental performance of three top-gate bilayer graphene FETs fabricated on SiC substrates from two different technologies, reported elsewhere \cite{YuHe17}, \cite{DenFad20}, \cite{YuHe17_}, has been described by the small-signal model proposed here. Technologies have been labeled here as technology 1 (T1) \cite{YuHe17}, \cite{YuHe17_} and technology 2 (T2) \cite{DenFad20}. T1 devices \cite{YuHe17}, \cite{YuHe17_} gate lengths $L_{\rm G}$s are of \SI{60}{\nano\meter} both of them and have gate widths $w_{\rm g}$s of ${\textstyle\SI{2}{}\times\SI{8}{\micro\meter}}$ \cite{YuHe17} and ${\textstyle\SI{2}{}\times\SI{15}{\micro\meter}}$ \cite{YuHe17_}, respectively. The device from T2 \cite{DenFad20} has $w_{\rm g}/L_{\rm g}$ of ${\textstyle\SI{2}{}\times\SI{15}{\micro\meter}/\SI{80}{\nano\meter}}$. Fabrication details can be found in \cite{FenYu14} for T1 \cite{YuHe17}, \cite{YuHe17_} and in \cite{FadWei18} for T2 \cite{DenFad20}. Table \ref{tab:param} lists the parameter values of the EC in Fig. \ref{fig:ckt} extracted here for each device as well as values reported in the corresponding references \cite{YuHe17}, \cite{DenFad20}, \cite{YuHe17_} where different topologies and effects than the ones included here have been considered, e.g., the contribution of the MG-interface related effects has not been included explicitly in the ECs associated to \cite{YuHe17}, \cite{DenFad20} and \cite{YuHe17_}.

\vspace{-0.3cm}
\begin{table} [!htb] 
\begin{center}
\caption{Reference and extracted values of small-signal EC shown in Fig. \ref{fig:ckt} for GFET technologies with different $w_{\rm{g}}/L_{\rm{g}} (\SI{}{\micro\meter}/\SI{}{\nano\meter})$.}
\begin{tabular}{c||c|c||c|c}

& \multicolumn{2}{c||}{Technology 1} & \multicolumn{2}{c}{Technology 2} \\ 

& \multicolumn{2}{c||}{\makecell{$\SI{16}{}/\SI{60}{}$ \cite{YuHe17} ($\SI{30}{}/\SI{60}{}$ \cite{YuHe17_})\\${\textstyle V_{\rm{GS}}=\SI{1.5}{\volt} (\SI{1}{\volt})}$,\\${\textstyle V_{\rm{DS}}=-\SI{0.5}{\volt} (-\SI{0.5}{\volt})}$}} & \multicolumn{2}{c}{\makecell{$\SI{30}{}/\SI{80}{}$ \cite{DenFad20}\\$V_{\rm{GS}}=\SI{1}{\volt}$,\\$V_{\rm{DS}}=\SI{1.5}{\volt}$}} \\ \hline \hline

parameter & \cite{YuHe17} (\cite{YuHe17_}) & this work & \cite{DenFad20} & this work  \\ \hline 

\multicolumn{5}{c}{intrinsic part} \\ \hline

$g_{\rm{m}} (\SI{}{\milli\siemens})$ & \SI{20}{} (\SI{28.5}{}) & \SI{20.3}{} (\SI{60}{}) & -\SI{3.8}{} & -\SI{10.7}{}  \\ 
$g_{\rm{ds}} (\SI{}{\milli\siemens})$ & \SI{70}{} (\SI{115}{}) & \SI{71.1}{} (\SI{242}{}) & \SI{31}{} & \SI{87.4}{}  \\ 
$C_{\rm{gd}} (\SI{}{\femto\farad})$ & \SI{4}{} (\SI{9}{}) & \SI{3.9}{} (\SI{7.8}{}) & \SI{4.4}{} & \SI{5.9}{}  \\
$C_{\rm{gs}} (\SI{}{\femto\farad})$ & \SI{13}{} (\SI{13}{}) & \SI{3.9}{} (\SI{7.8}{}) & \SI{9.7}{} & \SI{5.9}{} \\
$C_{\rm{ds}} (\SI{}{\femto\farad})$ & \SI{20}{} (\SI{20}{}) & \SI{20.7}{} (\SI{98.1}{}) & -- & \SI{21.4}{} \\ \hline

\multicolumn{5}{c}{contact resistances} \\ \hline

$R_{\rm{sc}} (\SI{}{\ohm})$ & \SI{5}{} (\SI{1.2}{}) & \SI{3.2}{} (\SI{1.9}{}) & -- & \SI{5.5}{}  \\
$R_{\rm{dc}} (\SI{}{\ohm})$ & \SI{7}{} (\SI{6.8}{})& \SI{4.5}{} (\SI{2.2}{}) & -- & \SI{16}{} \\ \hline

\multicolumn{5}{c}{extrinsic part} \\ \hline

$C_{\rm{gdp1}} (\SI{}{\femto\farad})$ & \SI{3}{} (\SI{1.8}{}) & \SI{3}{} (\SI{1}{}) & \SI{0.7}{} & \SI{0.7}{} \\ 
$C_{\rm{gdp2}} (\SI{}{\femto\farad})$ & -- (--) & \SI{1.8}{} (\SI{1.8}{}) & -- & \SI{0.5}{} \\ 
$C_{\rm{gsp1}} (\SI{}{\femto\farad})$ & \SI{6}{} (\SI{8.1}{}) & \SI{6}{} (\SI{8.1}{}) & \SI{8.2}{} & \SI{8.2}{} \\ 
$C_{\rm{gsp2}} (\SI{}{\femto\farad})$ & -- (--) & \SI{7.9}{} (\SI{8}{}) & -- & \SI{1.7}{}  \\ 
$C_{\rm{dsp1}} (\SI{}{\femto\farad})$ & \SI{12}{} (\SI{7.8}{}) & \SI{12}{} (\SI{7.8}{}) & \SI{7.8}{} & \SI{7.8}{} \\ 
$C_{\rm{dsp2}} (\SI{}{\femto\farad})$ & -- (--) & \SI{0}{} (\SI{0}{}) & -- & \SI{0}{} \\ 
$L_{\rm{ga}}   (\SI{}{\pico\henry})$ & \SI{33}{} (\SI{20}{}) & \SI{33}{} (\SI{20}{}) & \SI{60}{} & \SI{60}{}\\ 
$L_{\rm{sa}}   (\SI{}{\pico\henry})$ & \SI{9}{} (\SI{9}{}) & \SI{9}{} (\SI{9}{}) & \SI{0}{} & \SI{0}{} \\ 
$L_{\rm{da}}   (\SI{}{\pico\henry})$ & \SI{30}{} (\SI{18.7}{})& \SI{30}{} (\SI{18.7}{})  & \SI{40}{} & \SI{40}{}\\ 
$R_{\rm{ga}}   (\SI{}{\ohm})$ & \SI{2}{} (\SI{8.4}{})& \SI{2}{} (\SI{8.4}{}) & -- & \SI{57}{}\\
$R_{\rm{sa}} (\SI{}{\ohm})$ & -- (--)& \SI{1.8}{} (\SI{1.2}{}) & -- & \SI{0}{}\\
$R_{\rm{da}} (\SI{}{\ohm})$ & -- (--)& \SI{2.6}{} (\SI{6.8}{}) & -- & \SI{0}{}\\ 

\end{tabular} \label{tab:param}
\end{center}
\end{table} 

\vspace{-0.3cm}
The reference $R_{\rm{c}}$ values of the two T1 devices \cite{YuHe17}, obtained via cold-FET measurements \cite{YuHe17_}, have not been used here since it does not embrace the impact of lateral fields on the potential barrier at MG interfaces \cite{AnzMan18}. Instead, $R_{\rm{c}}$s for these T1 devices \cite{YuHe17}, \cite{YuHe17} have been extracted here  considering the impact of both lateral and vertical fields, $V_{\rm DS}$ and $V_{\rm GS}$, on the potential barrier with an extraction method presented elsewhere (see Eq. (1) in \cite{PacFei20}). This method \cite{PacFei20} yields a contact resistivity ${\textstyle R_{\rm{c}}\cdot w_{\rm g}}$ of \SI{123}{\ohm\cdot\micro\meter} for both devices from T1 \cite{YuHe17}, \cite{YuHe17_}. The approach adopted in this work implies lower $R_{\rm{sc}}$ and $R_{\rm{dc}}$ than in the reference ECs as well as the incorporation of access parasitic resistances. On the other hand, for the T2 device, the reference $R_{\rm c}$ value, obtained with the transfer length method (TLM), has not been explicitly included in the EC presented in \cite{DenFad20} and hence, the impact of the MG interface effects on the device has been embraced by other internal parameters. In contrast, such TLM-extracted $R_{\rm{c}}$, corresponding to a $R_{\rm{c}}\cdot w_{\rm g}$ of \SI{645}{\ohm\cdot\micro\meter}, has been used here in the proposed EC. The explicit incorporation of $R_{\rm{c}}$ and its non-symmetrical distribution in our approach yields different values of some of the extracted small-signal elements in comparison to the reported ones in both reference works.

Intrinsic parameters of the EC in Fig. \ref{fig:ckt} have been calculated from $[Y_{\rm i}]$  \cite{RamPac20} ($g_{m}={\rm Re}({y_{\rm 21,i}})$, $g_{\rm ds}={\rm Re}({y_{\rm 22,i}})$, $C_{\rm ds}={\rm Im}(y_{\rm 22,i})/\omega - C_{\rm gd}$ and $ C_{\rm gd/gs}=-{\rm Im}(y_{\rm 12,i})/\omega$) by using the $R_{\rm c}$ separation method (cf. Fig. \ref{fig:ckt_int}) towards a correct description of HF experimental data by the intrinsic device model and considering asymmetric contact resistances\footnote{The starting values to obtain this asymmetric disposal of $R_{\rm c}$ are the ones obtained with an \textit{I-V}-based extraction method for \cite{YuHe17} and with TLM for \cite{DenFad20}}. Available values of extrinsic parameters reported in the corresponding work (obtained by a de-embedding process discussed in Section II) have been used here while the remaining extrinsic parameter values have been obtained with a computer-aided least-square error-function \cite{RouEsc95} optimization towards a correct description of experimental raw $S$-parameters.

\section{Validation with experimental results} \label{ch:exp}

\begin{figure}[!htb]
\centering
\hspace{-0.15cm}
\includegraphics[height=0.32\textwidth]{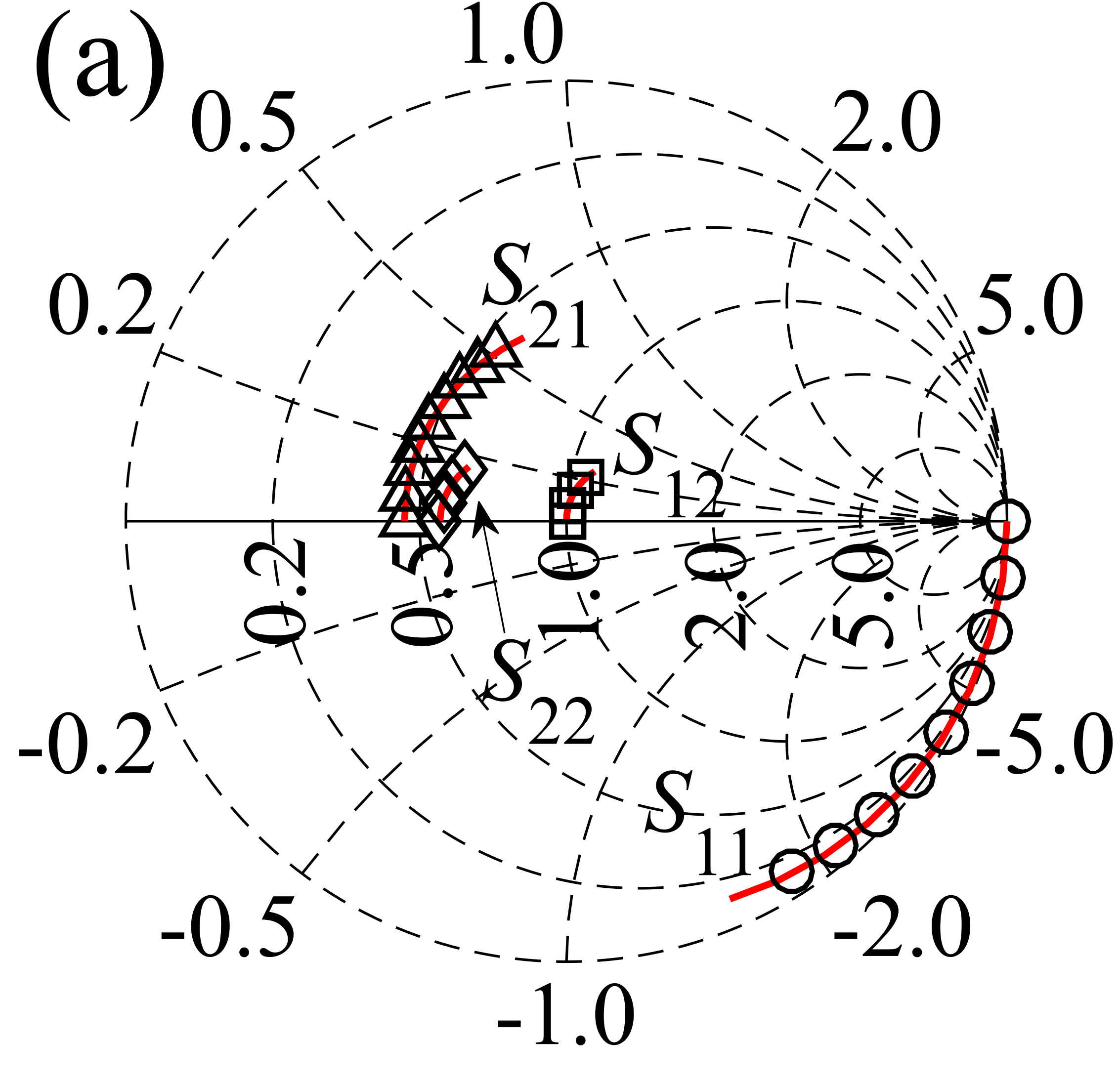} 
\includegraphics[height=0.32\textwidth]{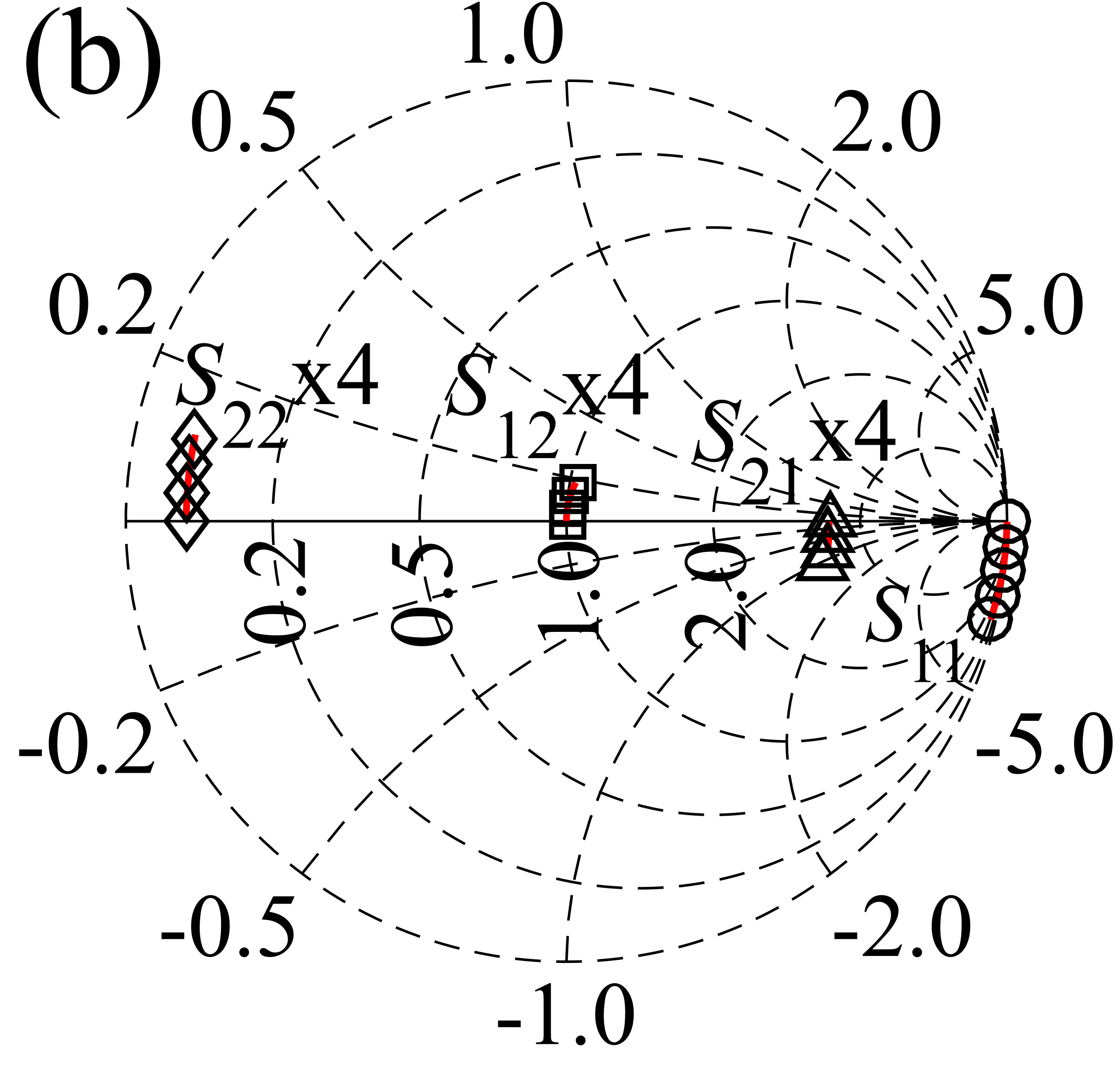} \\
\vspace{-0.3cm} \caption{$S$-parameters of (a) the smallest T1 GFET \cite{YuHe17} ${\textstyle(V_{\rm{GS}}=\SI{1.5}{\volt}}$, ${\textstyle V_{\rm{DS}}=-\SI{0.5}{\volt})}$ and (b) of the T2 GFET \cite{DenFad20} ${\textstyle(V_{\rm{GS}}=\SI{1}{\volt}}$, ${\textstyle V_{\rm{DS}}=\SI{1.5}{\volt})}$. Markers are experimental data and lines represent small-signal modeling results. Some data in (b) are increased 4 times for visualization purposes.}
\label{fig:Sxx}
\end{figure}

The small-signal model proposed here describes correctly the experimental $S$-parameters of the smallest GFET from T1 \cite{YuHe17} and the T2 device \cite{DenFad20} as shown in Fig. \ref{fig:Sxx}. 

\vspace{-0.2cm}

\begin{figure}[!htb]
\centering
\includegraphics[height=0.32\textwidth]{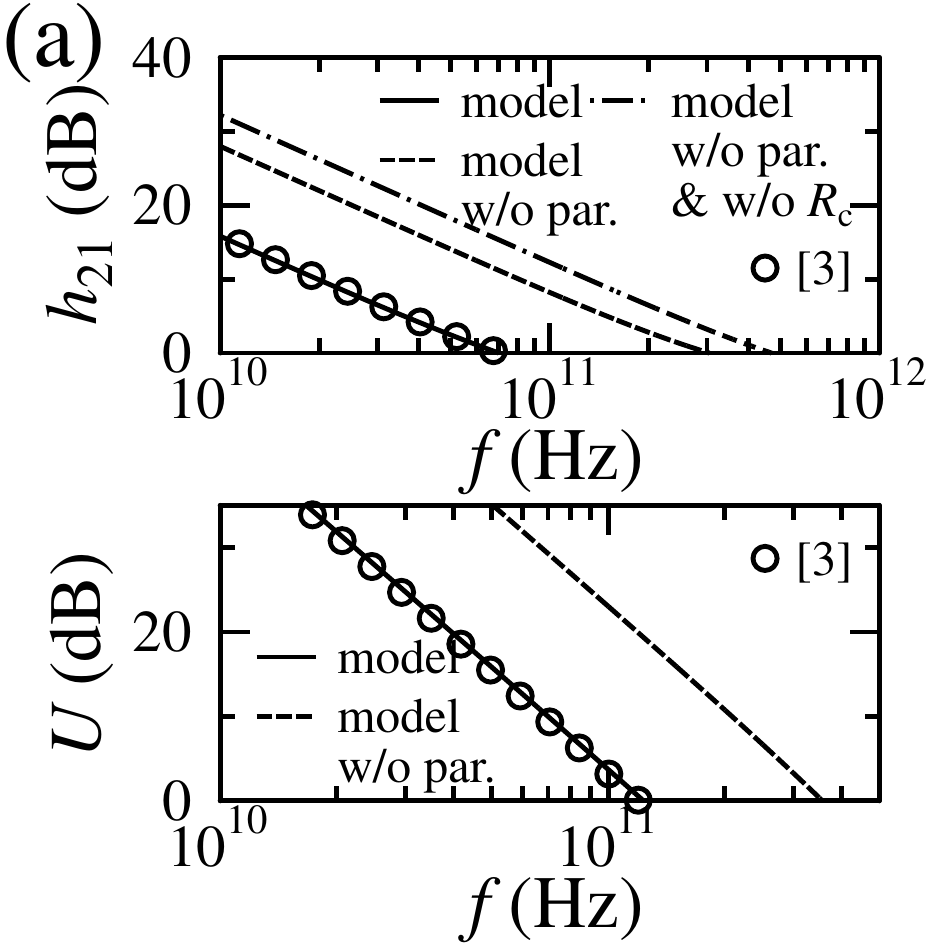}
\includegraphics[height=0.32\textwidth]{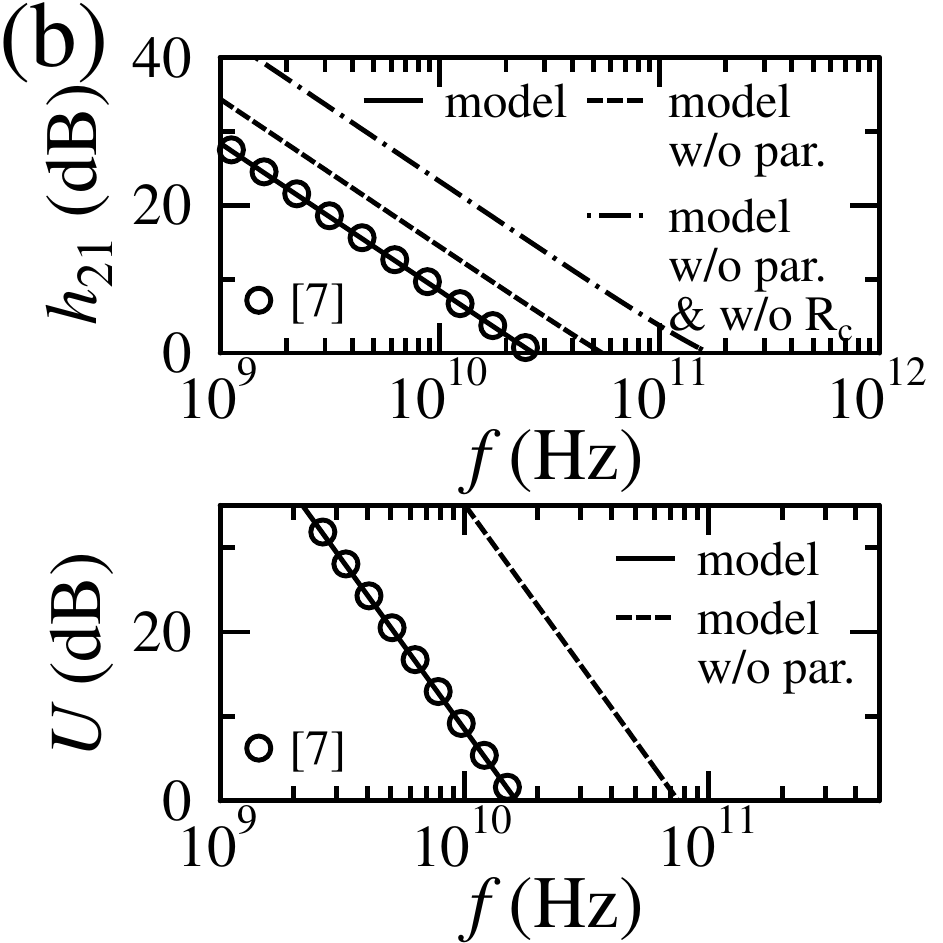} \\
\vspace{-0.3cm}\caption{Experimental (markers) and simulated (lines) $h_{21}$ (top) and $U$ (bottom) versus frequency for the (a) \SI{16}{\micro\meter}-wide/\SI{60}{\nano\meter}-long T1 device \cite{YuHe17}  ${\textstyle(V_{\rm{GS}}=\SI{1.5}{\volt}}$, ${\textstyle V_{\rm{DS}}=-\SI{0.5}{\volt})}$ and (b) for the T2 device (\SI{30}{\micro\meter}/\SI{80}{\nano\meter}) \cite{DenFad20} ${\textstyle(V_{\rm{GS}}=\SI{1}{\volt}}$, ${\textstyle V_{\rm{DS}}=\SI{1.5}{\volt})}$.}
\label{fig:h21_U}
\end{figure}

The experimental extrinsic transit frequency $f_{\rm{t,e}}$ and extrinsic maximum oscillation frequency $f_{\rm{max,e}}$ have been also described with the model proposed here as demonstrated in Fig. \ref{fig:h21_U} where the experimental and simulated small-signal current gain $h_{21}$ and the unilateral power gain $U$ versus frequency have been shown for each device. The EC proposed in \cite{YuHe17}, yielding a $f_{\rm{t,e}}$ and $f_{\rm{max,e}}$ of $\sim$\SI{72}{\giga\hertz} and $\sim$\SI{110}{\giga\hertz}, respectively,  is not capable to reproduce the corresponding experimental figures of merit (FoM) of \SI{70}{\giga\hertz} and \SI{120}{\giga\hertz}, respectively. In contrast, the approach considered here, including a more precise extraction and distribution of $R_{\rm c}$, and hence a more accurate description of small-signal elements, is a more efficient and reliable method to obtain an accurate HF modeling of such device as shown in Fig. \ref{fig:h21_U}(a).

Intrinsic FoM can be useful to indicate the HF material capabilities and performance projections for optimized technologies. In contrast to other studies where the impact of $R_{\rm{c}}$ is included indirectly in the intrinsic elements, the decomposition of the MG-related effects and of the internal transistor phenomena performed here (cf. Fig. \ref{fig:ckt_int}) enables the evaluation of the intrinsic cutoff frequency $f_{\rm t,i}$ with and without including the contribution of the contact resistances as shown in Figs. \ref{fig:h21_U}(a),(b). For the devices studied here, the lower the $R_{\rm{c}}$, the larger the difference between these $f_{\rm t,i}$s. E.g., the intrinsic $h_{21}$ without parasitics and $R_{\rm c}$ impact elucidates a superior performance of the bilayer graphene in \cite{YuHe17} over the one in \cite{DenFad20} which can be related to the different fabrication processes.

\begin{figure}[!htb]
\centering
\hspace{-0.15cm}
\includegraphics[height=0.32\textwidth]{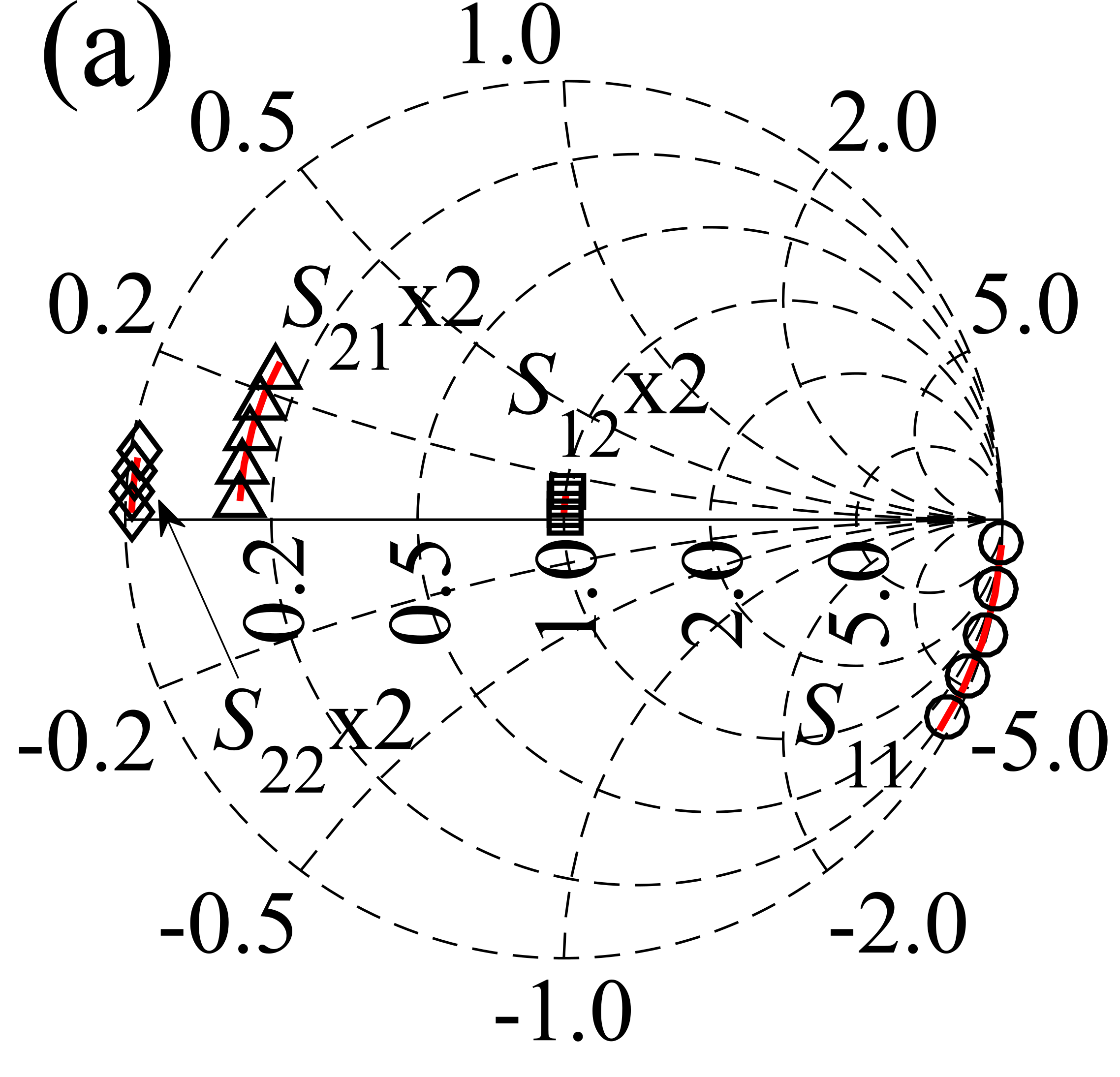} 
\includegraphics[height=0.32\textwidth]{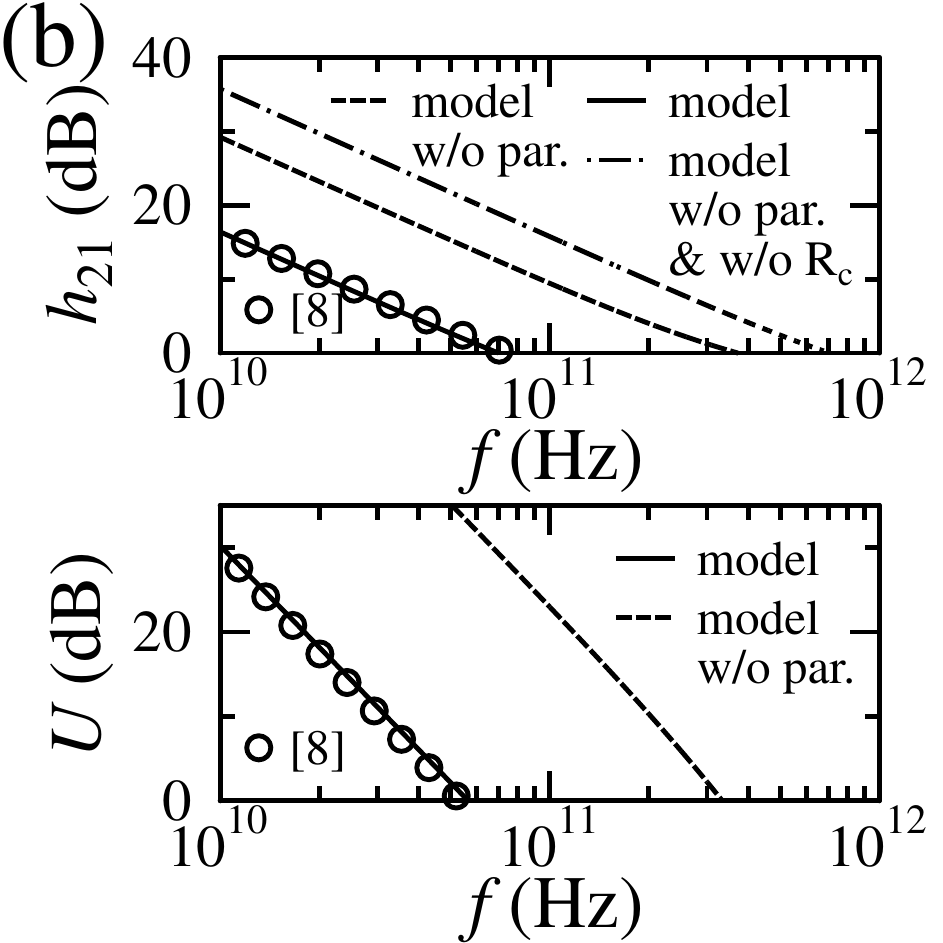} \\
\vspace{-0.3cm} \caption{Experimental (markers) and modeling results (lines) of the \SI{30}{\micro\meter}-wide/\SI{60}{\nano\meter}-long T1 device \cite{YuHe17_}: (a) $S$-parameters, (b) $h_{21}$ (top) and $U$ (bottom) versus frequency ${\textstyle(V_{\rm{GS}}=\SI{1}{\volt}}$, ${\textstyle V_{\rm{DS}}=-\SI{0.5}{\volt})}$.}
\label{fig:x}
\end{figure}

T1 devices \cite{YuHe17}, \cite{YuHe17_} are of different device footprint: $w_{\rm g}/L_{\rm G}=$\SI{16/60}{\micro\meter/\nano\meter} \cite{YuHe17} and \SI{30/60}{\micro\meter/\nano\meter} \cite{YuHe17_}. Some model parameters used for the correct HF description of the widest T1 device \cite{YuHe17_}, as shown in Fig. \ref{fig:x}, scale with $w_{\rm g}$. The same contact resistivity obtained for the smallest T1 device \cite{YuHe17} has been used for the widest one \cite{YuHe17_}, yielding an $R_{\rm c}$ of \SI{4.1}{\ohm} for the latter, i.e., $R_{\rm c}$ scales linearly with $w_{\rm g}$. Additionally, the intrinsic capacitances scale as $\sim$ $1/w_{\rm g}$ for this technology (see values in Table \ref{tab:param}). These scalability trends are enabled by similar devices material characteristics and a weak $V_{\rm GS}$-dependence of the electrostatics at channel edges (in the transport direction) \cite{FeiJim16} at the maximum $g_{\rm m}$-bias point in devices from this technology \cite{YuHe17}, \cite{YuHe17_}.

\vspace{-0.1cm}
\vspace{-0.3cm}

\section{Conclusion}

A small-signal GFET equivalent circuit with contact resistances embracing the effect of material layers and bias-dependent potential barriers associated to MG interfaces has been presented here. An extraction methodology enables to separate $R_{\rm c}$ effects from the intrinsic transistor and from the bias-independent de-embedded extrinsic elements. In contrast to other works, this approach enables a more exact quantification of intrinsic and extrinsic parts of the equivalent circuit. The proposed model has been verified with experimental $S$-parameters and HF FoM of three devices from two different GFET technologies. Intrinsic model parameters scale with the device footprint for one of the studied technologies. 


\begin{thebibliography}{1}

\bibitem{TanAnd14} M. Tanzid, M. A. Andersson, J. Sun, J. Stake, "Microwave noise characterization of graphene field effect transistors", \emph{Applied Physics Letters}, vol. 104, 013502, Jan. 2014. DOI: 10.1063/1.4861115


\bibitem{LyuLu16} H. Lyu, Q. Lu, J. Liu, X. Wu, J. Zhang, J. Li, J. Niu,
Z. Yu, H. Wu, H. Qian, "Deep-submicron graphene field-effect transistors with state-of-art $f_{\rm{max}}$", \emph{Scientific Reports}, vol. 6,  35717, Oct. 2016. DOI: 10.1038/srep35717

\bibitem{YuHe17} C. Yu, Z. Z. He, X. B. Song, Q. B. Liu, T. T. Han, S. B. Dun, J. J. Wang, C. J. Zhou, J. C. Guo, Y. J. Lv, Z. H. Feng, S. J. Cai, "Improvement of the Frequency Characteristics of Graphene Field-Effect Transistors on SiC Substrate", \emph{IEEE Electron Device Letters}, vol. 38, no. 9, pp. 1339-1342, Sep. 2017. DOI: 10.1109/LED.2017.2734938

\bibitem{PasWei17} F. Pasadas, W. Wei, E. Pallecchi, H. Happy, D. Jiménez, "Small-Signal Model for 2D-Material Based FETs Targeting Radio-Frequency Applications: The Importance of Considering Nonreciprocal Capacitances", \emph{IEEE Transactions on Electron Devices}, vol. 64, no. 11, pp. 4715-4716, Nov. 2017. DOI: 10.1109/TED.2017.2749503

\bibitem{SanXu18} L. Sang, Y. Xu, Y. Wu, R. Chen, "Device and Compact Circuit-Level Modeling of Graphene Field-Effect Transistors for RF and Microwave Applications", \emph{IEEE Transactions on Circuits and Systems I: Regular Papers}, vol. 65, no. 8, pp. 2559-2570, Aug. 2018. DOI: 10.1109/TCSI.2018.2793852

\bibitem{BonAsa19} M. Bonmann, M. Asad, X. Yang, A. Generalov, A. Vorobiev, L. Banszerus, C. Stampfer, M. Otto, D. Neumaier, J. Stake, "Graphene Field-Effect Transistors With High Extrinsic fT and fmax", \emph{IEEE Electron Device Letters}, vol. 40, no. 1, Jan. 2019. DOI: 10.1109/LED.2018.2884054

\bibitem{DenFad20} M. Deng, D. Fadil, W. Wei, E. Pallecchi, H. Happy,
G. Dambrine, M. De Matos, T. Zimmer, S. Fregonese, "High-Frequency Noise Characterization and Modeling of Graphene Field-Effect Transistors", \emph{IEEE Transactions on Microwave Theory and Techniques}, vol. 68, no. 6, pp. 2116-2123, Apr. 2020. DOI:  10.1109/TMTT.2020.2982396


\bibitem{YuHe17_} C. Yu, Z. Z. He, X. B. Song, Q. B. Liu, S. B. Dun, T. T. Han, J. J. Wang, C. J. Zhou, J. C. Guo, Y. J. Lv, S. J. Cai, Z. H. Fenga, "High-frequency noise characterization of graphene field effect transistors on SiC substrates", \emph{Applied Physics Letters}, vol. 111, 033502, Jul. 2017. DOI: 10.1063/1.4994324

\bibitem{GiuDiB17} F. Giubileo, A. Di Bartolomeo, "The role of contact resistance in graphene field-effect devices", \emph{Progress in Surface Science}, vol. 92, pp. 143-175, Jun. 2017. DOI: 10.1016/j.progsurf.2017.05.002

\bibitem{HsuWan11} A. Hsu, H. Wang, K. K. Kim, J. Kong, T. Palacios, "Impact of graphene interface quality on contact resistance and RF device performance", \emph{IEEE Electron Device Letters}, vol. 32, no. 8, pp. 1008-1010, Aug., 2011. DOI: 10.1109/LED.2011.2155024

\bibitem{ChaJim15} F. A. Chaves, D. Jiménez, A. A. Sagade, W. Kim, J. Riikonen, H. Lipsanen, D. Neumaier, "A physics-based model of gate-tunable metal–graphene contact resistance benchmarked against experimental data", \emph{2D Materials}, vol. 2, 025006, May 2015. DOI: 10.1088/2053-1583/2/2/025006

\bibitem{PacFei20} A. Pacheco-Sanchez, P. C. Feijoo, D. Jimenez, "Contact resistance extraction of graphene FET technologies based on individual device characterization", \emph{Solid-State Electronics}, vol. 172, 107882, Oct. 2020. DOI: 10.1016/j.sse.2020.107882

\bibitem{RamPac20} J. N. Ramos-Silva, A. Pacheco-Sánchez, E. Ramírez-Garcia, D. Jiménez "Small-signal parameters extraction and noise analysis of CNTFETs", \emph{Semiconductor Science and Technology}, vol. 35, 045024, Mar. 2020. DOI: 10.1088/1361-6641/ab760b

\bibitem{TieHav05} L. F. Tiemeijer, R. J. Havens, A. B. M. Jansman, Y. Bouttement, "Comparison of the “Pad-Open-Short” and “Open-Short-Load” Deembedding Techniques for Accurate On-Wafer RF Characterization
of High-Quality Passives", \emph{IEEE Transactions on Microwave Theory and Techniques}, vol. 53, no. 2, pp. 723-729, Feb. 2005. DOI: 10.1109/TMTT.2004.840621

\bibitem{PasJim20} F. Pasadas, D. Jiménez, "Non-Quasi-Static Effects in Graphene Field-Effect Transistors Under High-Frequency Operation", \emph{IEEE Transactions on Electron Devices}, vol. 67, no. 5, pp. 2188-2196, Apr. 2020. DOI: 10.1109/TED.2020.2982840

\bibitem{FenYu14} Z. H. Feng. C. Yu, J. Li, Q. B. Liu, Z. Z. He, X. B. Song, J. J. Wang, S. J. Cai, "An ultra clean self-aligned process for highmaximum oscillation frequency graphene transistors", \emph{Carbon}, vol. 75, pp. 249-254, Aug. 2014. DOI: 10.1016/j.carbon.2014.03.060

\bibitem{FadWei18} D. Fadil, W. Wei, M. Deng, S. Fregonese, W. Strupinski, E. Pallecchi, H. Happy, "2D-Graphene Epitaxy on SiC for RF Application: Fabrication, Electrical Characterization and Noise Performance", in Proc. \emph{IEEE/MTT-S International Microwave Symposium (IMS)}, Philadelphia, USA, Jun. 2018. DOI: 10.1109/MWSYM.2018.8439655 

\bibitem{RouEsc95} J. P. Roux, L. Escotte, R. Plana, J. Graffeuil, S. L. Delage, H. Blanck, "Small-signal and noise model extraction technique for heterojunction bipolar transistor at microwave frequencies", \emph{IEEE Transactions on Microwave Theory and Techniques}, vol. 43, no. 2, pp. 293-298, 1995. DOI: 10.1109/22.348087

\bibitem{AnzMan18} L. Anzi, A. Mansouri, P. Pedrinazzi, E. Guerriero, M. Fiocco, A. Pesquera, A. Centeno, A. Zurutuza, A. Behnam, E. A. Carrion, E. Pop, R. Sordan, "Ultra-low contact resistance in graphene devices at the Dirac point", \emph{2D Materials}, vol. 5, 025014, Feb. 2018. DOI: 10.1088/2053-1583/aaab96

\bibitem{FeiJim16} P. C. Feijoo, D. Jiménez, X. Cartoixà, "Short channel effects in graphene-based field effect transistors targeting radio-frequency applications", \emph{2D Materials}, vol. 3, 025036, Jun. 2016. DOI: 10.1088/2053-1583/3/2/025036


\end{thebibliography}

\end{document}